\documentclass[aps,superscriptadress,preprint]{revtex4}
\usepackage{graphicx}
\usepackage{mathrsfs}
\usepackage{hyperref}

\begin{document}
\title{Different Regular Black Holes: Geodesic Structures of Test Particles}

\author{Zihan Xi$^{1}$} 
\author{Chen Wu$^{2}$}
\author{Wenjun Guo$^{1,}$\footnote{Electronic address: impgwj@126.com}}
\affiliation{
\small 1.University of Shanghai for Science and Technology, Shanghai 200093,China \\2. Shanghai Institute of Applied Physics, Chinese Academy of
Sciences, Shanghai 201800, China}

\begin{abstract}
This paper investigates the metric of previously proposed regular black holes, calculates their effective potentials, and plots the curves of the effective potentials. By determining the conserved quantities, the dynamical equations for particles and photons near the black hole are derived. The analysis encompasses timelike and null geodesics in different spacetimes, including bound geodesics, unstable circular geodesics, stable circular geodesics, and escape geodesics. The findings are presented through figures and tables. Furthermore, the bound geodesics of the four regular black hole spacetimes are analyzed, examining the average distance of particle orbits from the center of the event horizon, the precession behavior of the perihelion, and the probability of particles appearing inside the outer event horizon during motion. Based on these analyses, a general formula is proposed, which yields the existing metrics when specific parameter values are chosen. The impact of parameter variations on the effective potential and geodesics is then computed using this new formula.
\end{abstract}

 \maketitle

\section{Introduction}
Black holes are a significant astronomical entity that holds great research value. Accurately determining their properties through meticulous calculations and thorough observations is crucial for advancing our understanding of these phenomena. N. Heidari has proposed a new analytical method for computing the quasinormal modes of black holes by employing the Rosen-Morse potential to estimate the quasi-normal frequencies of Schwarzschild black holes. By performing numerical calculations and comparisons, the authors have demonstrated that this approach outperforms previous techniques in terms of accuracy\cite{1}.Moreover, this method offers a valuable contribution to our understanding of the physical properties of black holes and promotes further advancements in related research areas. Chen Wu's investigation of gravitational perturbations and quasinormal mode (QNM) frequencies around some regular black holes suggests that the Wentzel-Kramers-Brillouin (WKB) approximation and asymptotic iteration method can be used to perform a detailed analysis of the frequencies of gravitational QNMs. His research results indicate that the imaginary part of the quasinormal frequencies as a function of the charge parameter exhibits different monotonic behaviors for different black hole spacetimes. Moreover, the article provides an asymptotic expression for gravitational QNMs using the Eikonal limit method and proves the stability of gravitational perturbations in these spacetimes\cite{2}. The findings of this research article significantly enhance our understanding of the properties of the gravitational field and the stability of black holes. Sometimes breakthroughs in other fields can also advance our understanding of black holes. After the discovery of gravitational waves in 2015, researchers found that by analyzing the gravitational wave signals, they could determine some characteristics of black holes that were previously unproven, such as the high mass of black hole binaries and the near-absence of spin in black hole binaries. The combination of gravitational waves and black hole research is a popular research direction in the study of black holes \cite{3,4,5}. In order to obtain more precise fundamental properties of black holes, it is crucial to identify more reliable methods. Zening Yan employed three distinct methods in their investigation of Schwarzschild-Tangherlini black hole spacetime and numerically validated that the third-order WKB approximation outperformed higher-order WKB approximations in their study\cite{6}. Such research on methods can be of great assistance to future researchers.

As the ultimate properties of a black hole are uniquely determined by its mass, charge, and angular momentum, researchers often choose to study charged black holes\cite{7}, or rotating black holes \cite{8}, or black holes surrounded by other matter\cite{9}. Mubasher Jamila investigated the dynamics of particles around a Schwarzschild-like black hole surrounded by dark energy and an external magnetic field. They found that regardless of the charge of the test particles, the radius of their innermost stable circular orbit (ISCO) and their orbital frequency were strongly influenced by the magnetic field\cite{10}. These findings have important theoretical and practical implications for furthering our understanding of black hole physics and noncommutative geometry, among other fields. Studying multiple types of black holes is therefore crucial as the properties of different black hole spacetimes vary significantly. Yen-Kheng Lim investigated the geodesic equations of charged and uncharged particles in the Ernst metric and found that their orbits can only be stable when the electric field strength is below a certain critical value\cite{11}.

In the study of black holes, investigating geodesics is particularly important. Sheng Zhou and others investigated the geodesic structures of test particles in the Bardeen spacetime. By analyzing the effective potential, they identified the timelike and null geodesic trajectories in the Bardeen spacetime, and described the possible orbits of particles and photons using diagrams\cite{12}.E. Kapsabelis investigated the geodesics of the Schwarzschild-Finsler-Randers (SFR) spacetime and compared their model with the corresponding model in general relativity. They found small differences in the deflection angles between SFR spacetime and general relativity, which can be attributed to the anisotropic metric structure of the model and the Randers term\cite{13}.Jiri Podolsky investigated some properties of the extreme Schwarzschild-de Sitter spacetime. By studying geodesics and the deviation equation of geodesics, they Obtained the conclusion that a specific group of observers can escape the singularity of the black hole. This paper proposed the synchronous coordinates system in the paper, which provides a basis for the further development of Black Hole No-Hair Theory\cite{14}.

\section{Regular black hole and Orbit equation}
\subsection{Regular black hole}
The general line element representing spherically symmetric regular BH is given by
\begin{equation}
ds^2= -f(r)dt^2 + f(r)^{-1}dr^2 + r^2(d\theta ^2 + \sin^2\theta d\phi ^2 ),
\end{equation}
Where $t$,$r$,$\theta$,$\phi$ represent ordinary spacetime spherical coordinates, and the lapse function $f(r)$ is determined by the specific spacetime. Chi Zhang conducted an investigation on several regular black hole spacetimes\cite{15}.This article selects the following regular black hole spacetimes. The new regular black hole spacetime discovered by Hayward in 2006 Hayward\cite{16},This black hole is similar to the one discovered by Bardin\cite{17}.Its characteristic is that as the radius tends to infinity, it rapidly approaches zero, indicating that it does not exert a significant gravitational influence on the surrounding space in regions far from the event horizon. The lapse function is
\begin{equation}
f(r)=\left(1 - \frac{2Mr^2}{r^2+2\alpha^2}\right)
\end{equation}
Where $\alpha$ is assumed to be a positive constant and M represents the mass of the black hole. The number of event horizons can be 0, 1, or 2, depending on the relative values of $M$ and $\alpha$.

Eloy Ayón-Beato proposed a regular exact black hole solution, which is a charged black hole with a source that satisfies the weak energy condition in nonlinear electrodynamics\cite{18}.The lapse function for this black hole is given as
\begin{equation}
	f(r)=\left(1-{2Mr^2\over (r^2+q^2)^{2\over 3}} + \frac{q^2r^2}{(r^2+q^2)^2}\right)
\end{equation}
Where q represents the charge.

Bronnikov introduced nonlinear electrodynamics, considering the Born-Infeld theory, and constructed a regular black hole\cite{19,20},The lapse function for this black hole is given as
\begin{equation}
	f(r)=\left(1 - {2M\over r}(1-\tanh{r_0\over r}\right)
\end{equation}
Where $r_0$ is related to the electric charge. Afterwards, following the work of Bronnikov\cite{19,20},Dymnikova established a regular spherically symmetric charged black hole\cite{21}.He considered the coupling of nonlinear electrodynamics with general relativity. The lapse function for Dymnikova’s solution is given as
\begin{equation}
f(r)=\left(1-{4M\over\pi r}\left(\arctan{r\over r_0} - {rr_0\over r^2+r_0^2}\right)\right)
\end{equation}
Where $r_0=\frac{\pi}{8}\frac{q^2}{M}$, it is a length scale defined, and $q$ represents the charge.

The black holes selected in this study have been organized into Table \ref{t1}, with the parameters used for each black hole provided.
\begin{table}[hbt]\centering\caption{The black holes selected in this paper}
	\begin{tabular*}{16.5cm}{*{4}{c @{\extracolsep\fill}}}
		\hline \hline
		Lapse function & Extremal condition  & Reference & Originator    \\ \hline
		
		$f=\left(1-{2Mr^2\over r^3+2\alpha^2}\right)$ & $\alpha \approx 1.06 $ &  \cite{16}  & Hayward       \\
		
		$f=\left( 1-{2Mr^2\over (r^2+q^2)^{3\over 2}}+{q^2r^2\over (r^2+q^2)^2}\right)$ &$q \approx 0.63 $  & \cite{18}  & Ay\'on-Beato and Garc\'ia \\
		
		$f=\left(1-{2M\over r}\left(1- \mbox{tanh}{r_0\over r}\right)\right)$ &$r_0 \approx 0.55$ &  \cite{19,20}  & Bronnikov  \\
		
		$f=\left(1-{4M\over \pi r}\left(\arctan{r\over r_0}- {rr_0\over r^2+r_0^2}\right)\right)$ & $r_0 = 0.45$ & \cite{21}  &Dymnikova  \\  \hline \hline
	\end{tabular*} \label{t1}
\end{table}

\subsection{Orbit equation}

The general line element representing spherically symmetric regular black holes is given by $(1)$, and the corresponding Lagrangian can be obtained from the variational principle as:
\begin{equation}
\mathcal{L}=\frac{1}{2}\left[-f\left(r\right){\dot{t}}^2+f\left(r\right)^{-1}{\dot{r}}^2+r^2\left({\dot{\theta}}^2+\sin^2{\theta}{\dot{\varphi}}^2\right)\right]	
\end{equation}
For photons, $L=0$. For particles, by choosing $\lambda$ as $\tau$, $L=\frac{1}{2}$. Define
\begin{equation}
 \eta=\Biggl\{\begin{array}{cc}
 	0&(photons)
 	\\1&(particles)
 \end{array}
\end{equation}
Then, the Lagrangian $\mathcal{L}=\frac{\eta}{2}$. Choosing $\tau$ as the affine parameter, the Lagrangian equation is 

\begin{equation}
	\frac{d}{d\tau}\frac{\partial\mathcal{L}}{\partial{\dot{x}}^\nu}-\frac{\partial\mathcal{L}}{\partial x^\nu}=0
\end{equation}

Since the metric is static and spherically symmetric, it is not a function of time $t$ and azimuthal angle $\varphi$, yielding
\begin{equation}
	\frac{\partial\mathcal{L}}{\partial\dot{t}}=-f\left(r\right)\dot{t}=-E
\end{equation}	
\begin{equation}
	\frac{\partial\mathcal{L}}{\partial\dot{\varphi}}=r^2\sin^2{\theta}\dot{\varphi}=L
\end{equation}	
Where $E$ and $L$ are two conserved quantities. If we choose the initial condition as $\theta=\frac{\pi}{2}$, then we have $\dot{\theta}=0$ , $\ddot{\theta}=0$. In this case, we can derive the orbital equation for the equatorial plane selected in this paper.
\begin{equation}
	{\dot{r}}^2=E^2-f\left(r\right)\left(\eta+\frac{L^2}{r^2}\right)
\end{equation}
we define $f\left(r\right)\left(\eta+\frac{L^2}{r^2}\right)$ as the effective potential $V_{eff}^2$, and equation $(11)$ can be rewritten as ${\dot{r}}^2=E^2-V_{eff}^2$. Making the substitution $r=\frac{1}{u}$, the orbital equation can be transformed into
\begin{equation}
	\left(\frac{du}{d\varphi}\right)^2=\frac{E^2}{L^2}-\frac{f\left(\frac{1}{u}\right)\eta}{L^2}-f\left(\frac{1}{u}\right)u^2
\end{equation}

\section{ The geodesic structure of different spacetimes}
\subsection{Hayward spacetime}

When $\eta=1$, it corresponds to a timelike geodesics, and its effective potential is
\begin{equation}
	V_{eff}^2=\left(1-\frac{2Mr^2}{r^3+2\alpha^2}\right)\left(1+\frac{L^2}{r^2}\right)
\end{equation}
The orbital equation for the particle is
\begin{equation}
	\left(\frac{du}{d\varphi}\right)^2=\frac{E^2-\eta}{L^2}+\frac{2\eta Mu+2ML^2u^3}{L^2\left(1+2\alpha^2u^3\right)}-u^2
\end{equation}
Upon further differentiation, the second-order orbital equation can be obtained
\begin{equation}
	\frac{d^2u}{d\varphi^2}=-\frac{m\left(4\alpha^3u^3-1\right)+L^2u\left[\left(2\alpha^2u^3+1\right)^2-3mu\right]}{\left(2\alpha^2Lu^3+L\right)^2}
\end{equation}
Numerical analysis of the equation reveals the types of orbits and provides insights into how changes in the parameters of the Hayward black hole spacetime affect the geodesic structure.

As shown in Fig.1,$E_{\uppercase\expandafter{\romannumeral1}}^2$ and $E_{\uppercase\expandafter{\romannumeral2}}^2$represent critical energy values. Depending on the particle's energy satisfying different conditions, three different types of motion orbits can be observed, namely bound orbits, circular orbits, and escape orbits.

In the case of bound orbits: when $E_{\uppercase\expandafter{\romannumeral1}}^2<E^2<E_{\uppercase\expandafter{\romannumeral2}}^2$, the effective potential curve indicates the presence of two types of bound orbits, as shown in Fig.2, at this energy level.

(1) The particle is confined in a bound orbit within the range $r_A<r<r_B$, where $r_A$ and $r_B$ represent the pericenter and apocenter of the planetary orbit, respectively. This orbit has a self-intersection point and exhibits counterclockwise precession.

(2) The particles are bound in a bound orbit within the range of $r_C<r<r_D$, where the distance between $r_C$ and $r_D$ is much larger compared to (1). The orbit has one self-intersection point and exhibits a greater degree of clockwise precession compared to (1).

In the case of circular orbits: as shown in Fig.3 and Fig.4, at critical energy levels, two types of circular orbits can be observed. For higher critical energy, the orbit is an unstable circular orbit, while for lower critical energy, the orbit is a stable circular orbit. 

(1) As depicted in Fig.3, when $E=E_{\uppercase\expandafter{\romannumeral2}}$ , it orbits on an unstable circular orbit with $r=r_B$. Any slight perturbation causes the orbit to transition into two other types of orbits, where the particle is bound between $r_A<r<r_B$ or the particle is bound between $r_B<r<r_C$.

(2) As illustrated in Fig.4 , when $E=E_{\uppercase\expandafter{\romannumeral1}}$, the particle's motion follows a stable circular orbit. Even in this case, there are still two scenarios. In one scenario, the particle moves on a bound orbit, confined between $r_A<r<r_B$, and the orbit exhibits self-intersection points. In another scenario, the particle moves on a stable circular orbit at $r=r_C$. 

In the case of escape orbits : when the particle's energy $E$ is greater than the critical energy $E_{\uppercase\expandafter{\romannumeral2}}$, three different types of escape orbits occur. As depicted in Fig.5, when $E=E_A^2$, the particle's orbit is a curved escape orbit. The particle follows a curved path from infinity towards the vicinity of the black hole, experiences deflection along the orbit, forms self-intersection points with the previous orbit, and then returns to infinity. When $E=E_B^2$, the particle's orbit does not curve, and deflection only occurs near the central region. After forming self-intersection points, the particle follows a straight-line path back to infinity. When $E=E_C^2$, the particle's orbit is a straight-line orbit without self-intersection points. After deflection, the particle returns to infinity along a straight path.

\subsection{Ayón-Beato and García spacetime}

In the Ayón-Beato and García spacetime, when $\eta=1$, it corresponds to timelike geodesics, and the effective potential is given by
\begin{equation}
	V_{eff}^2=\left(1-\frac{2Mr^2}{\left(r^2+q^2\right)^\frac{3}{2}}+\frac{q^2r^2}{\left(r^2+q^2\right)^2}\right)\left(1+\frac{L^2}{r^2}\right)
\end{equation}
When the energy equals $E$, the bound orbit exhibits a significantly greater precession of the particle's orbit pericenter compared to the bound orbit in the Hayward spacetime, as illustrated in Fig.6 .Its circular and escape orbits resemble those in the Hayward spacetime.When the energy $E$ is equal to $E_{\uppercase\expandafter{\romannumeral2}}$, small perturbations cause changes in the particle's orbit. When the energy $E$ is equal to $E_{\uppercase\expandafter{\romannumeral1}}$, the particle's orbit is either a bound orbit or a stable circular orbit. For $E$ exceeding $E_{\uppercase\expandafter{\romannumeral2}}$, the particle's orbit becomes an escape orbit.

\subsection{Bronnikov spacetime}
In Bronnikov spacetime, when $\eta=1$, the trajectory corresponds to timelike geodesics, with the effective potential denoted as 
\begin{equation}
	V_{eff}^2=\left(1-\frac{2M}{r}\left(1-\tanh{\frac{r_0}{r}}\right)\right)\left(1+\frac{L^2}{r^2}\right)
\end{equation}

The effective potential can be illustrated as depicted in Figure 7, wherein the presence of three types of orbits persists: bound orbit, circular orbit, and escape orbit. The specific details of these orbits are presented in Table \ref{t2}.

\begin{table}[hbt]\centering\caption{The category of null geodesic types in the Bronnikov spacetime with $r_0=0.55$,$L=3.5$,$M=1$ ,$E_{\uppercase\expandafter{\romannumeral1}}^2=0.910$,$E_{\uppercase\expandafter{\romannumeral2}}^2=1.11$}
	\begin{tabular*}{16.5cm}{*{4}{c @{\extracolsep\fill}}}
		\hline \hline
	  &	Energy & The situation of geodesics  &      \\ \hline
		
		  & $E^2=E_{\uppercase\expandafter{\romannumeral1}}^2$ & Bound orbit and stable circle orbit &        \\
		
	 &	$E_{\uppercase\expandafter{\romannumeral1}}^2<E^2<1$ &Two types of bounded orbits &  \\
		
	 &	$1<E^2<E_{\uppercase\expandafter{\romannumeral2}}^2$ &Bound orbit and escape orbit &  \\
		
	 &	$E_{\uppercase\expandafter{\romannumeral2}}^2<E^2$ & Escape orbit &  \\  \hline \hline
	\end{tabular*} \label{t2}
\end{table}

When $\eta=0$, it corresponds to null geodesics, characterized by an effective potential denoted as 
\begin{equation}
	V_{eff}^2=\left(1-\frac{2M}{r}\left(1-\tanh{\frac{r_0}{r}}\right)\right)
\end{equation}
As shown in Fig.8, when $E^2=E_{\uppercase\expandafter{\romannumeral2}}^2$, the particle's orbit is an unstable circular orbit. Perturbations can alter the orbit's configuration. However, in contrast to the particle's behavior, photons are confined between $r_A$ and $r_B$, or the photon orbit is a circular orbit at $r=r_B$. Then, the curvature of the orbit gradually decreases, extending to infinity. When $E_{\uppercase\expandafter{\romannumeral2}}^2<E^2$, as depicted in Fig.9, the photon's orbit becomes an escape orbit. When $E^2=E_A^2$, photons curve near the black hole, experiencing deflection upon entering the event horizon. After exiting the event horizon, the photon's orbit forms two self-intersection points before returning along the curve to infinity. As the energy increases, the photon's orbit becomes a straight line with only one self-intersection point. With further energy increase, the self-intersection points vanish, and the photon enters along a straight line, experiencing deflection near the outer event horizon, and ultimately returning along a straight path to infinity.

\subsection{Dymnikova spacetime}
In the Dymnikova spacetime, the effective potential is denoted as
\begin{equation}
	V_{eff}^2=\left(1-\frac{4M}{\pi r}\left(\arctan{\frac{r}{r_0}}{-\frac{rr_0}{r^2+r_0^2}}\right)\right)\left(\eta+\frac{L^2}{r^2}\right)
\end{equation}
As illustrated in Fig.10, when $\eta=1$ in the Dymnikova spacetime, it corresponds to timelike geodesics. The orbital behavior is similar to that in the Bronnikov spacetime. Depending on the energy level, there can exist bounded orbits, stable circular orbits, and escape orbits. When $\eta=0$, it corresponds to null geodesics, and the specific orbital characteristics are depicted in Table \ref{t3} .

\begin{table}[hbt]\centering\caption{The category of null geodesic types in the Dymnikova spacetime with $r_0=0.55$,$L=3.5$,$M=1$,$E_{\uppercase\expandafter{\romannumeral2}}^2=0.86$}
	\begin{tabular*}{16.5cm}{*{4}{c @{\extracolsep\fill}}}
		\hline \hline
		&	Energy & The situation of geodesics  &      \\ \hline
		
		& $0<E^2<E_{\uppercase\expandafter{\romannumeral2}}^2$ & Bound orbit and escape orbit &        \\
		
		&	$E^2=E_{\uppercase\expandafter{\romannumeral2}}^2$ &Unstable circular bound orbit and unstable circular escape orbit &  \\
		
		&	$E_{\uppercase\expandafter{\romannumeral2}}^2<E^2$ &Escape orbit &  \\
		
		  \hline \hline
	\end{tabular*} \label{t3}
\end{table}

As depicted in Fig.11, when $\eta=1$ and the energy is taken at an intermediate value between the two extrema, six types of timelike bound geodesics can be obtained. By performing calculations, the average distance of the particle orbit from the center of the event horizon, the precession of the perihelion of the orbit, and the probability of the particle appearing inside the outer event horizon during motion can be determined. The specific data is presented in Table \ref{t4}. From this table, we can observe that for cases (\uppercase\expandafter{\romannumeral1}), (\uppercase\expandafter{\romannumeral3}), (\uppercase\expandafter{\romannumeral5}), and (\uppercase\expandafter{\romannumeral6}), the difference between the average distance from the center of the event horizon and the midpoint between the perihelion $r_A$ and aphelion $r_B$, namely $\frac{1}{2}\left(r_A+r_B\right)$, is small. This suggests that the particle orbits are relatively evenly distributed between the perihelion and aphelion. On the other hand, cases (\uppercase\expandafter{\romannumeral2}) and (\uppercase\expandafter{\romannumeral4}) exhibit an average distance closer to the perihelion, indicating a denser distribution of orbits in the vicinity of the perihelion. Furthermore, it can be observed that cases (\uppercase\expandafter{\romannumeral1}) and (\uppercase\expandafter{\romannumeral2}) exhibit counterclockwise precession, while the others exhibit clockwise precession. Additionally, case (\uppercase\expandafter{\romannumeral3}) demonstrates the smallest precession angle of the perihelion, while case (\uppercase\expandafter{\romannumeral4}) exhibits the largest precession angle. This observation may be related to the properties of the Ayón-Beato and García spacetime. The table also reveals that the probability of the particle orbit being inside the outer event horizon is smaller compared to it being outside the outer event horizon. This implies that particles are more likely to move outside the event horizon. 

\begin{table}[hbt]\centering\caption{The six types of timelike bound geodesics, their respective average distances from the center of the event horizon, precession angles of the perihelion, and probabilities of the orbit being inside the outer event horizon.}
	\begin{tabular*}{16.5cm}{*{5}{c @{\extracolsep\fill}}}
		\hline \hline
	Geodesics	&	Originator & Average distance  & Precession angle &  Probability    \\ \hline
		
	(\uppercase\expandafter{\romannumeral1})	& Hayward spacetime & 2.436 &  Counterclockwise $0.23\pi$ &         \\
		
	(\uppercase\expandafter{\romannumeral2})	& Hayward spacetime & 7.524 &  Counterclockwise $0.31\pi$ &         \\
		
	(\uppercase\expandafter{\romannumeral3})	& Ayón-Beato and García spacetime & 1.509 &  Clockwise $0.08\pi$ & 34.27\%        \\
	
	(\uppercase\expandafter{\romannumeral4})	& Ayón-Beato and García spacetime & 10.269 &  Clockwise $0.6\pi$ & 0\%        \\
		
	(\uppercase\expandafter{\romannumeral5})	& Bronnikov spacetime & 1.200 &  Clockwise $0.46\pi$ &  37.27\%       \\
			
	(\uppercase\expandafter{\romannumeral6})	& Dymnikova spacetime & 1.101 &  Clockwise $0.25\pi$ &  37.97\%       \\
				
		\hline \hline
	\end{tabular*} \label{t4}
\end{table}

\subsection{New metric formula}

After conducting research on the aforementioned four spacetimes, this paper concludes with a derived formula for the effective potential
\begin{equation}
	V_{eff}^2=\left(1-\frac{\alpha r^2}{\left(r^2+\beta\right)^\frac{3}{2}}+\frac{\beta r^2}{\left(r^2+\beta\right)^2}\right)\left(\eta+\frac{L^2}{r^2}\right)
\end{equation}

The parameter $\alpha$ is associated with the mass $M$, while $\beta$ is related to the charge $q$. Fig.12 presents the plots of the effective potential for different values of $\alpha$, ranging from 1.6 to 2.7 in increments of 0.1, with $\beta$ fixed at 0.3 and 0.35. It can be observed that as $\alpha$ increases, the number of extremal points in the effective potential gradually reduces from three to one.

By varying the parameters of the formula, the effective potential for the four previously selected spacetimes in this study can be obtained. The specific results are presented in Table \ref{t5}.

\begin{table}[hbt]\centering\caption{The values of the parameters in the formula for each spacetime}
	\begin{tabular*}{16.5cm}{*{3}{c @{\extracolsep\fill}}}
		\hline \hline
		$\beta$	&	$\alpha$ & Originator    \\ \hline
		
		0.3	& 1.664 & Ayón-Beato and García spacetime         \\
		
		0.3	& 1.737 & Bronnikov spacetime         \\
		
		0.3	& 1.771 & Dymnikova spacetime         \\
		
		0.3	& 2.391 & Hayward spacetime         \\
		
		0.35	& 1.838 & Ayón-Beato and García spacetime         \\
		
		0.35	& 1.917 & Bronnikov spacetime         \\
		
		0.35	& 1.953 & Dymnikova spacetime         \\
		
		0.35	& 2.637 & Hayward spacetime         \\
		\hline \hline
	\end{tabular*} \label{t5}
\end{table}

Table \ref{t6} presents the relationship between the values of $\alpha$ and the number of types of bound orbits when $\beta$ is set to 0.3 or 0.35, and the energy is taken at an intermediate value between the two extrema. 

\begin{table}[hbt]\centering\caption{The relationship between the values of $\alpha$ and $\beta$ and the number of types of bound orbits.}
	\begin{tabular*}{16.5cm}{*{3}{c @{\extracolsep\fill}}}
		\hline \hline
		$\beta$	&	$\alpha$ & The number of types of bound orbits    \\ \hline
		
		0.3	& $1.6\le\alpha<1.634$ & One         \\
		
		0.3	& $1.634\le\alpha<2.170$ & Two         \\
		
		0.3	& $2.170\le\alpha\le2.7$ & One         \\
		
		0.35	& $1.6\le\alpha<1.745$ & One         \\
		
		0.35	& $1.745\le\alpha<2.196$ & Two         \\
		
		0.35	& $2.196\le\alpha\le2.7$ & One         \\
		
		\hline \hline
	\end{tabular*} \label{t6}
\end{table}

When $\beta$ is set to 0.3 or 0.35, the bound geodesics for $\alpha$ values ranging from the minimum to the maximum, with the energy taken at an intermediate value, are depicted in Fig.12. By performing calculations, the precession angles of the perihelion for each bound geodesics can be obtained. The specific data is presented in Table \ref{t7}.

\begin{table}[hbt]\centering\caption{The precession angles of the perihelion for bound geodesics with continuous variations of $\alpha$, when $\beta$ is set to 0.3 and 0.35}
	\begin{tabular*}{16.5cm}{*{4}{c @{\extracolsep\fill}}}
		\hline \hline
	Timelike bound geodesics &	$\beta$	&	$\alpha$ & Precession angle of the perihelion    \\ \hline
		
	(\uppercase\expandafter{\romannumeral1}) & 0.3	& 1.8 & 0.49         \\
		
	(\uppercase\expandafter{\romannumeral1}) & 0.3	& 1.9 & 0.27         \\
	
	(\uppercase\expandafter{\romannumeral1}) & 0.3	& 2.0 & 0.11         \\
	
	(\uppercase\expandafter{\romannumeral1}) & 0.3	& 2.1 & 0.99         \\
	
	(\uppercase\expandafter{\romannumeral1}) & 0.35	& 1.8 & 0.51         \\
	
	(\uppercase\expandafter{\romannumeral1}) & 0.35	& 1.9 & 0.32         \\
	
	(\uppercase\expandafter{\romannumeral1}) & 0.35	& 2.0 & 0         \\
	
	(\uppercase\expandafter{\romannumeral1}) & 0.35	& 2.1 & 0.67         \\
		
		\hline \hline
	\end{tabular*} \label{t7}
\end{table}

\section{CONCLUSION}

This paper investigates a selection of regular black hole spacetimes\cite{15}, specifically focusing on four regular black hole spacetimes. Starting from the general line element of spherically symmetric regular black holes, the corresponding Lagrangian is derived using the variational principle. After determining the conserved quantities, setting $\theta=\frac{\pi}{2}$ allows the derivation of the orbital equations in the equatorial plane. By further differentiating these equations, the second-order orbital equations are obtained. By analyzing the orbital equations, this paper identifies different types of orbits in all cases, including bound orbits, stable circular orbits, unstable circular orbits, and escape orbits. The cases of $\eta=1$ and $\eta=0$ are discussed separately for both timelike and null geodesics in different spacetimes. The research focuses on the analysis of bound orbits in the aforementioned four regular black hole spacetimes. Through analyzing and calculating particle orbit data, the average distance of the particle orbit from the center of the event horizon is determined. By analyzing the periodicity of the orbits, the precession behavior of the perihelion is obtained. Finally, by analyzing the particle's position during motion, the probability of the particle appearing inside the outer event horizon is determined. In conclusion, this paper presents a new metric formula. When specific values of the parameters $\alpha$ and $\beta$ are chosen, existing metrics can be obtained. The impact of parameter variations of $\alpha$ and $\beta$ on the effective potential and geodesics is calculated and analyzed.

\newpage

\begin{figure}[tbp]
\includegraphics[width=13cm,height=11cm]{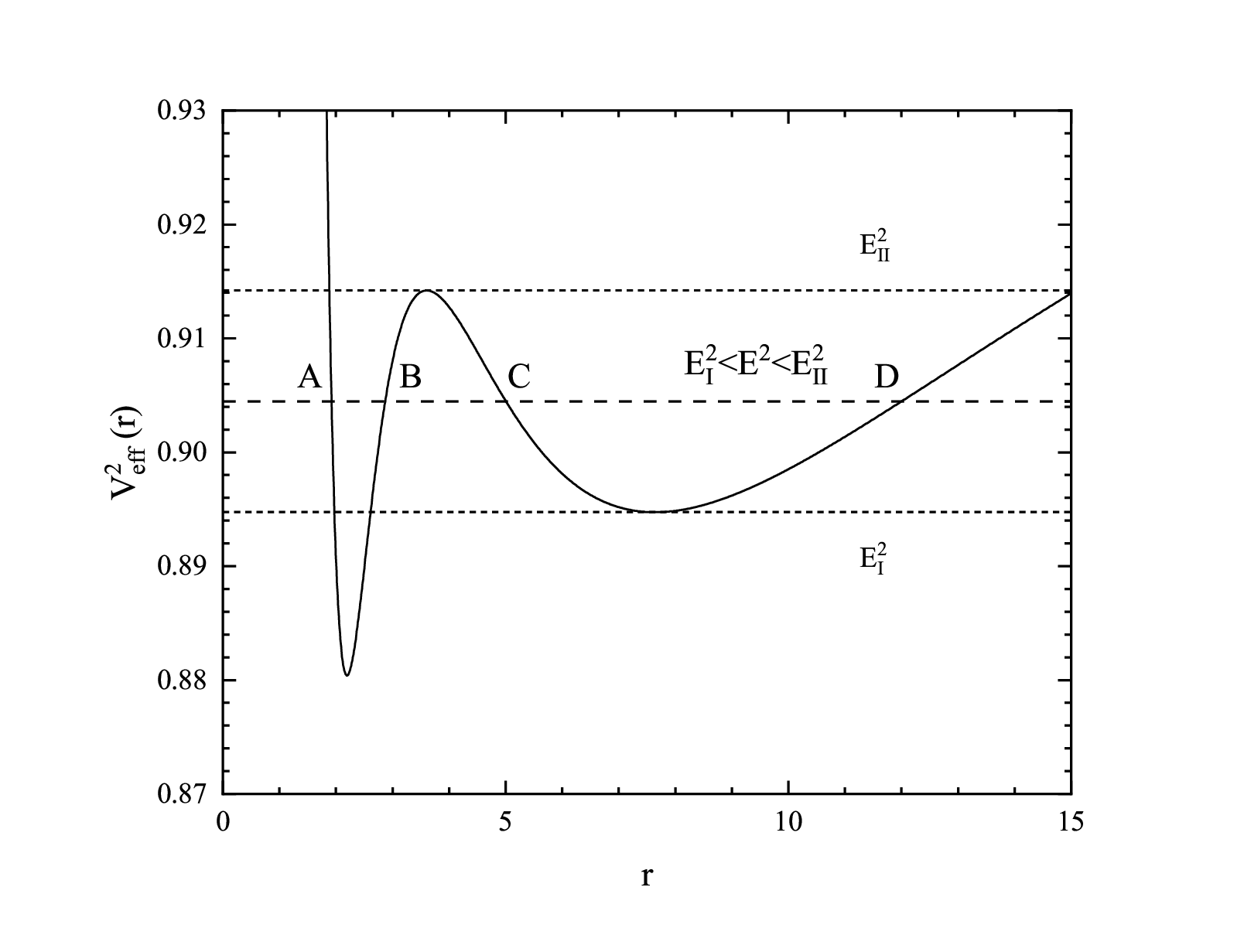}
\caption{The effective potential of timelike bound geodesics in the Hayward spacetime with $\alpha=1.06$, $L=3.5$, $M=1$, $E^2=0.904$, $E_{\uppercase\expandafter{\romannumeral1}}^2=0.895$, $E_{\uppercase\expandafter{\romannumeral2}}^2=0.914$}
\end{figure}

\begin{figure}[tbp]
\includegraphics[width=10.4cm,height=20cm]{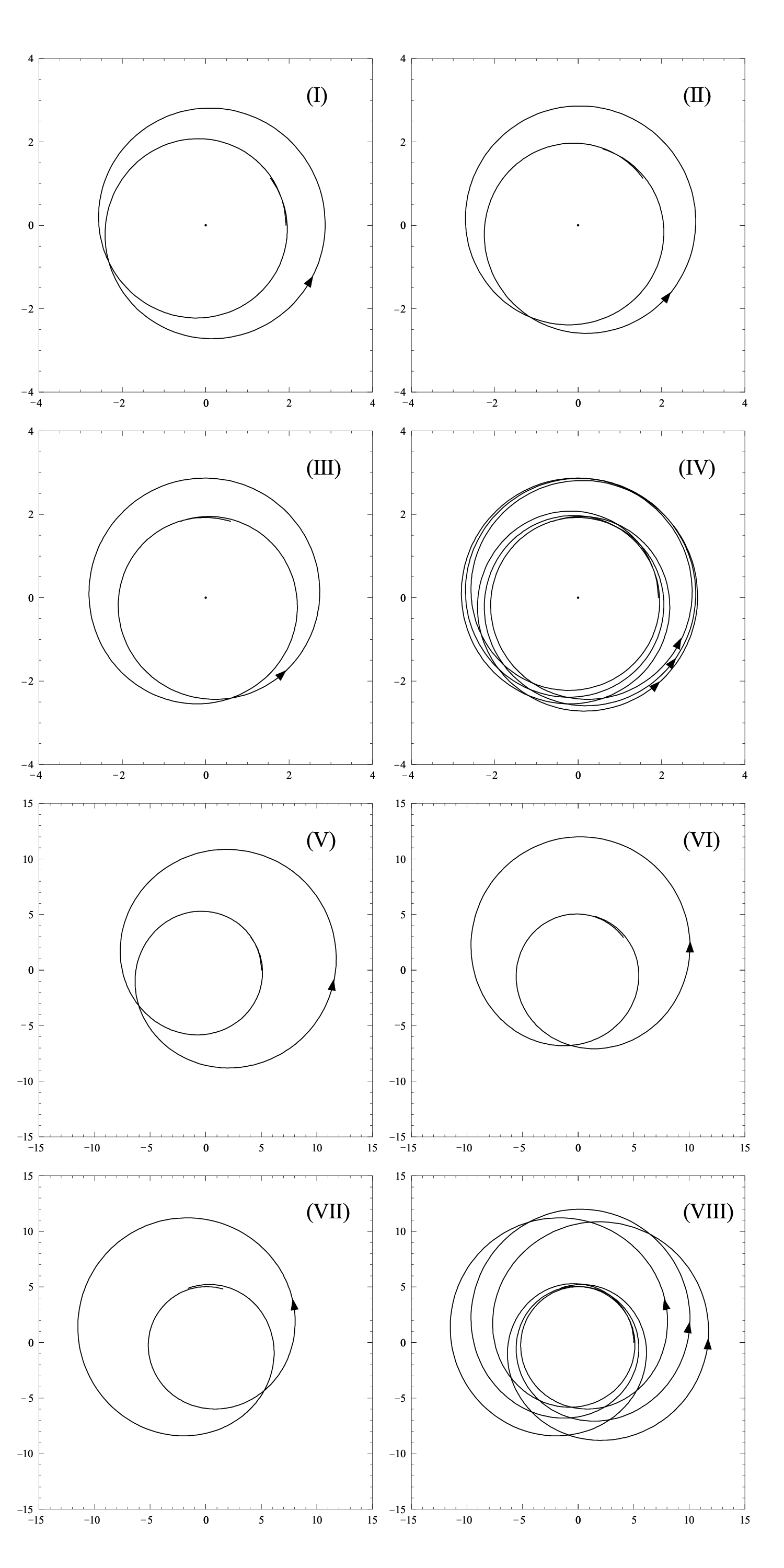}
\caption{The timelike bound geodesics in the Hayward spacetime with $\alpha=1.06$, $L=3.5$, $M=1$, $E^2=0.904$ [(\uppercase\expandafter{\romannumeral1}) + (\uppercase\expandafter{\romannumeral2}) + (\uppercase\expandafter{\romannumeral3}) = (\uppercase\expandafter{\romannumeral4}),(\uppercase\expandafter{\romannumeral5}) + (\uppercase\expandafter{\romannumeral6}) + (\uppercase\expandafter{\romannumeral7}) = (\uppercase\expandafter{\romannumeral8})]. }
\end{figure}

\begin{figure}[tbp]
\includegraphics[width=13cm,height=4.5cm]{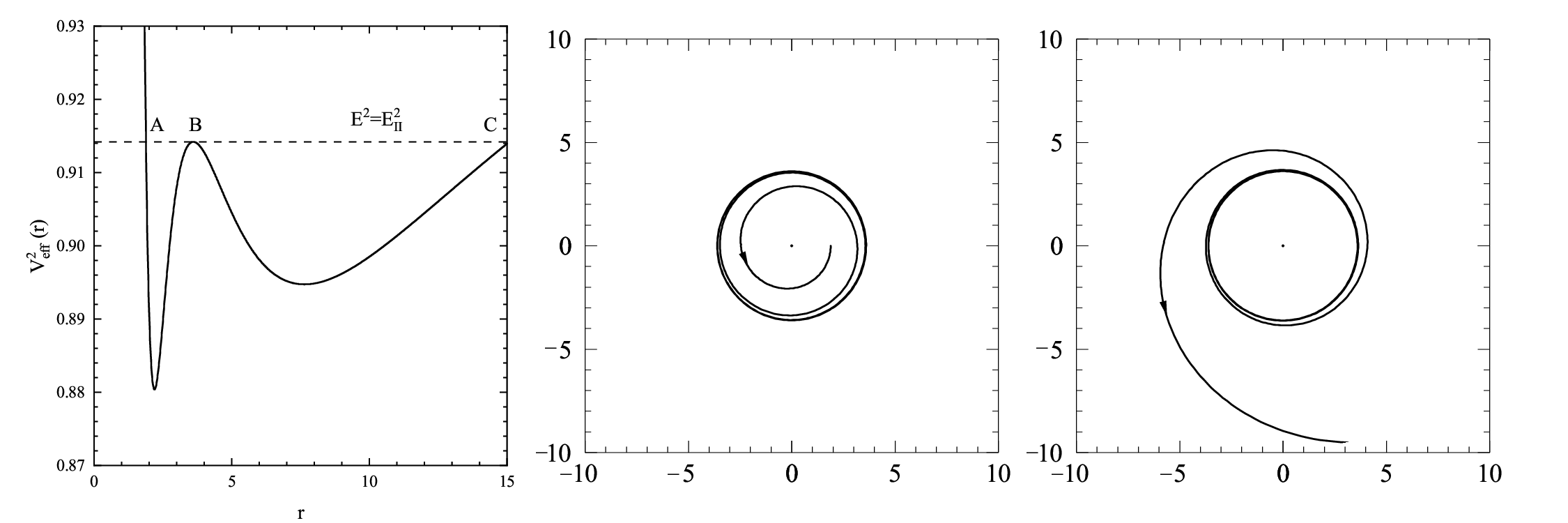}
\caption{ The unstable timelike circle geodesics in the Hayward spacetime with $\alpha=1.06$, $L=3.5$, $M=1$, $E^2=0.914$  }
\end{figure}

\begin{figure}[tbp]
	\includegraphics[width=13cm,height=4.5cm]{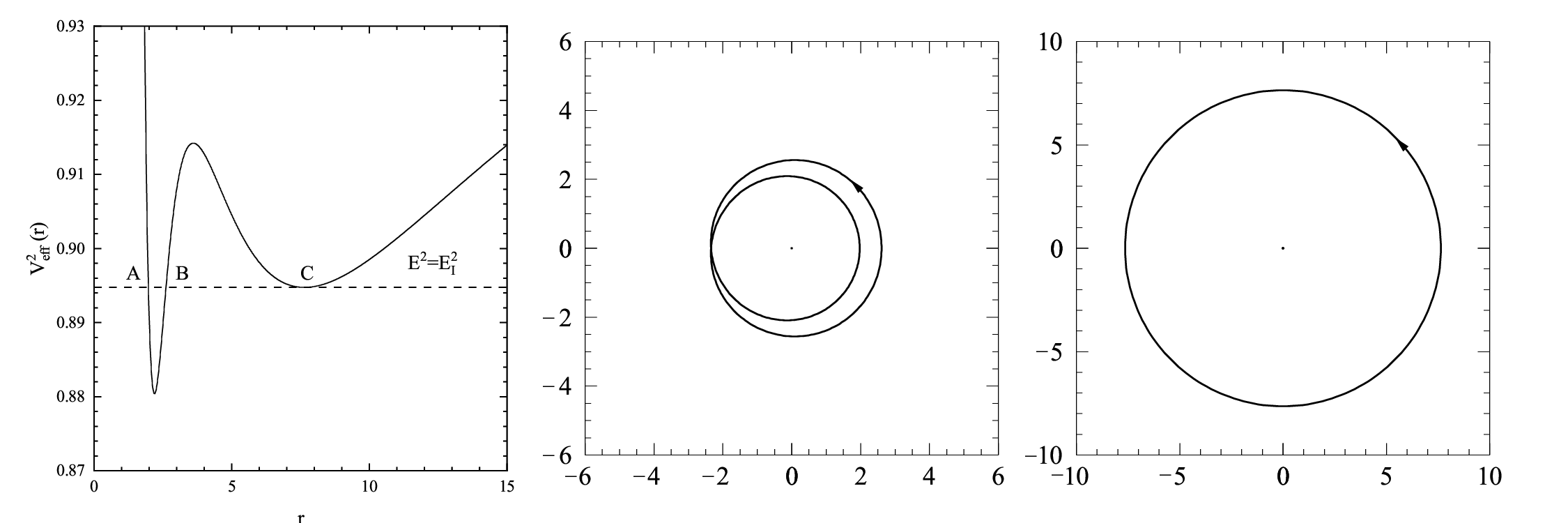}
	\caption{ The stable timelike circle geodesics in the Hayward spacetime with $\alpha=1.06$, $L=3.5$, $M=1$, $E^2=0.895$  }
\end{figure}

\begin{figure}[tbp]
	\includegraphics[width=13cm,height=13cm]{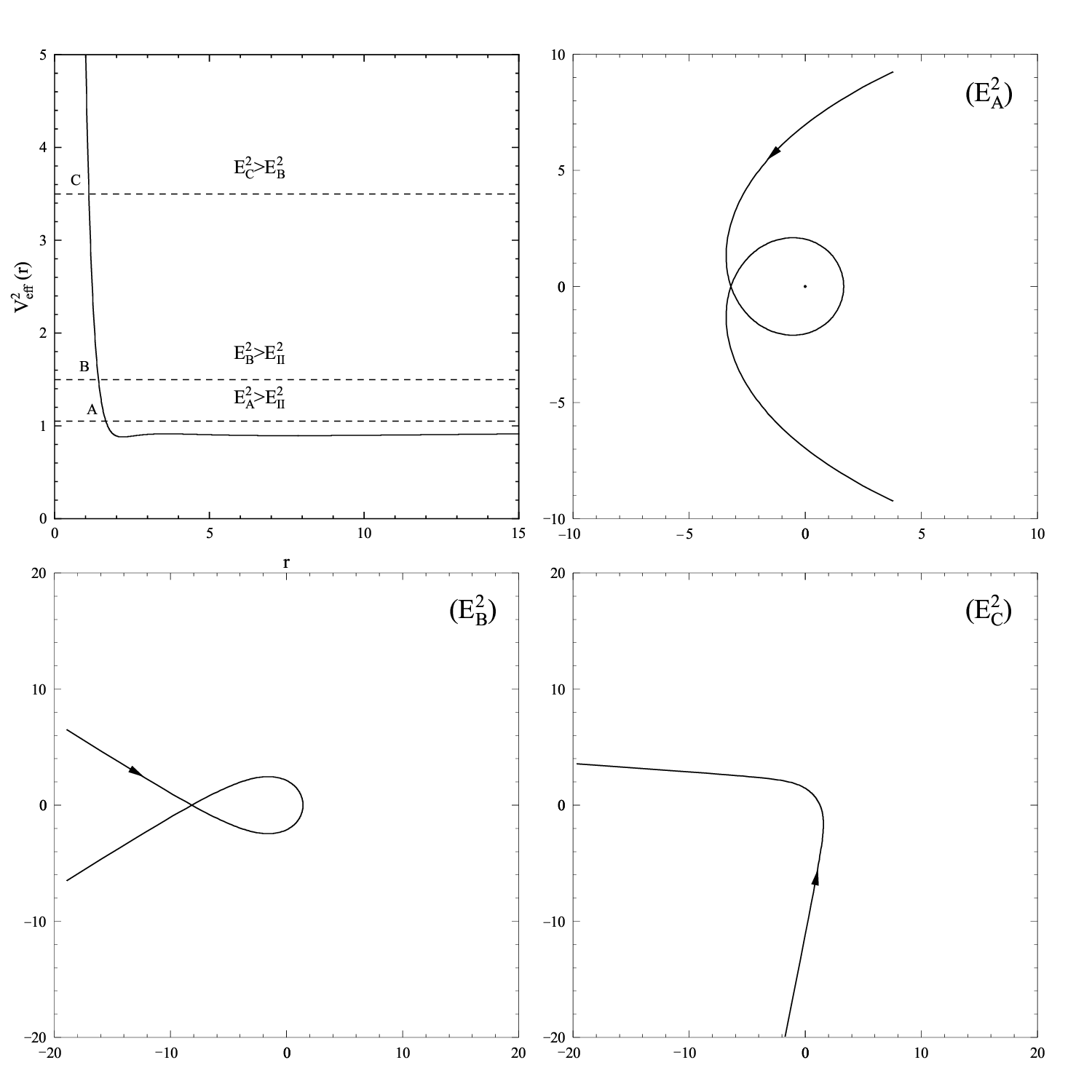}
	\caption{ The timelike escape geodesics in the Hayward spacetime with $\alpha=1.06$, $L=3.5$, $M=1$, $E_{\uppercase\expandafter{\romannumeral2}}^2=0.914$, $E_A^2=1.05$, $E_B^2=1.5$,$E_C^2=3.5$  }
\end{figure}

\begin{figure}[tbp]
	\includegraphics[width=13cm,height=6cm]{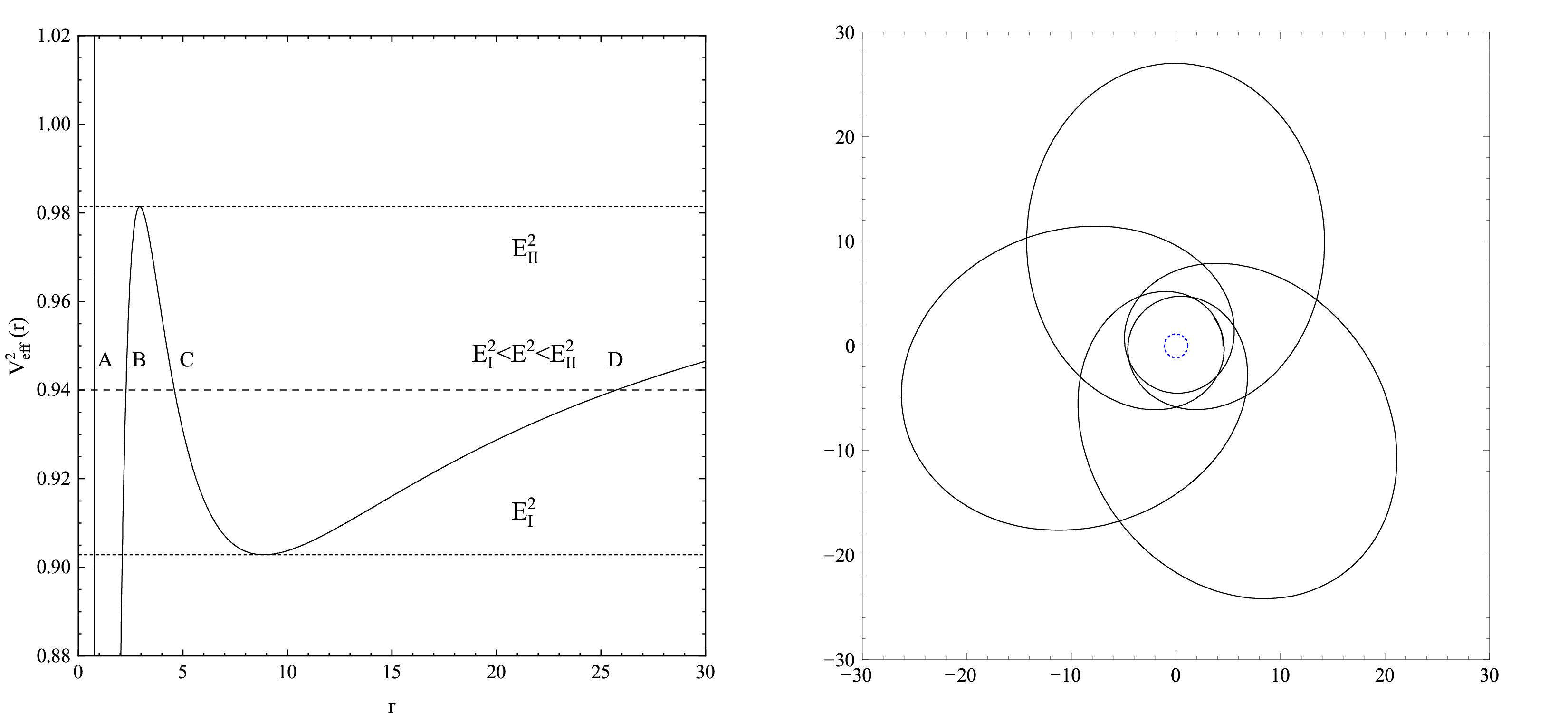}
	\caption{ The effective potential of timelike bound geodesics and the timelike bound geodesics in the Ayón-Beato and García spacetime with $q=0.63$, $L=3.5$, $M=1$, $E^2=0.94$,$E_{\uppercase\expandafter{\romannumeral1}}^2=0.90$, $E_{\uppercase\expandafter{\romannumeral2}}^2=0.98$  }
\end{figure}

\begin{figure}[tbp]
	\includegraphics[width=13cm,height=10cm]{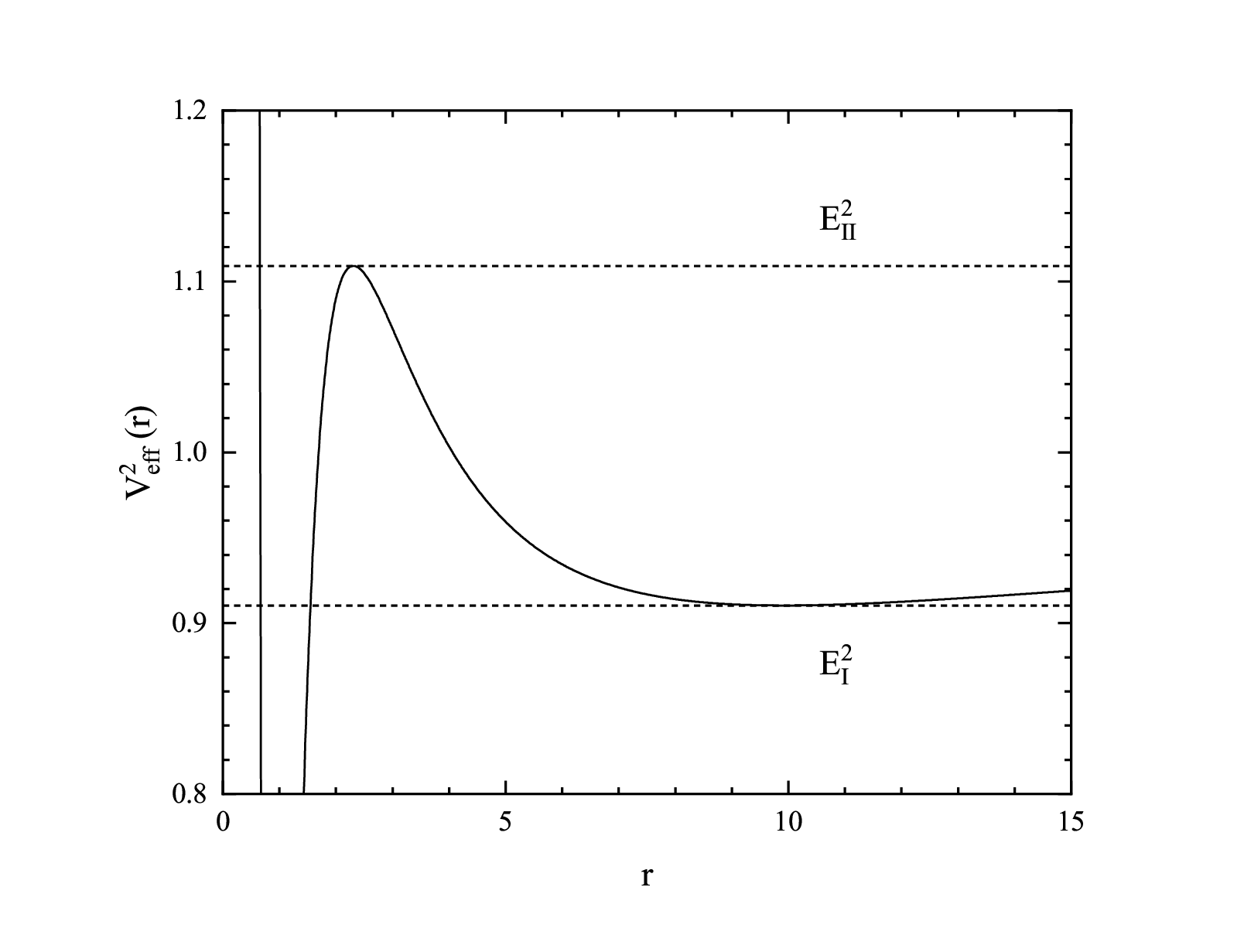}
	\caption{ The effective potential of timelike bound geodesics in the Bronnikov spacetime with $r_0=0.55$, $L=3.5$, $M=1$ , $E_{\uppercase\expandafter{\romannumeral1}}^2=0.910$, $E_{\uppercase\expandafter{\romannumeral2}}^2=1.11$ }
\end{figure}

\begin{figure}[tbp]
	\includegraphics[width=13cm,height=4.5cm]{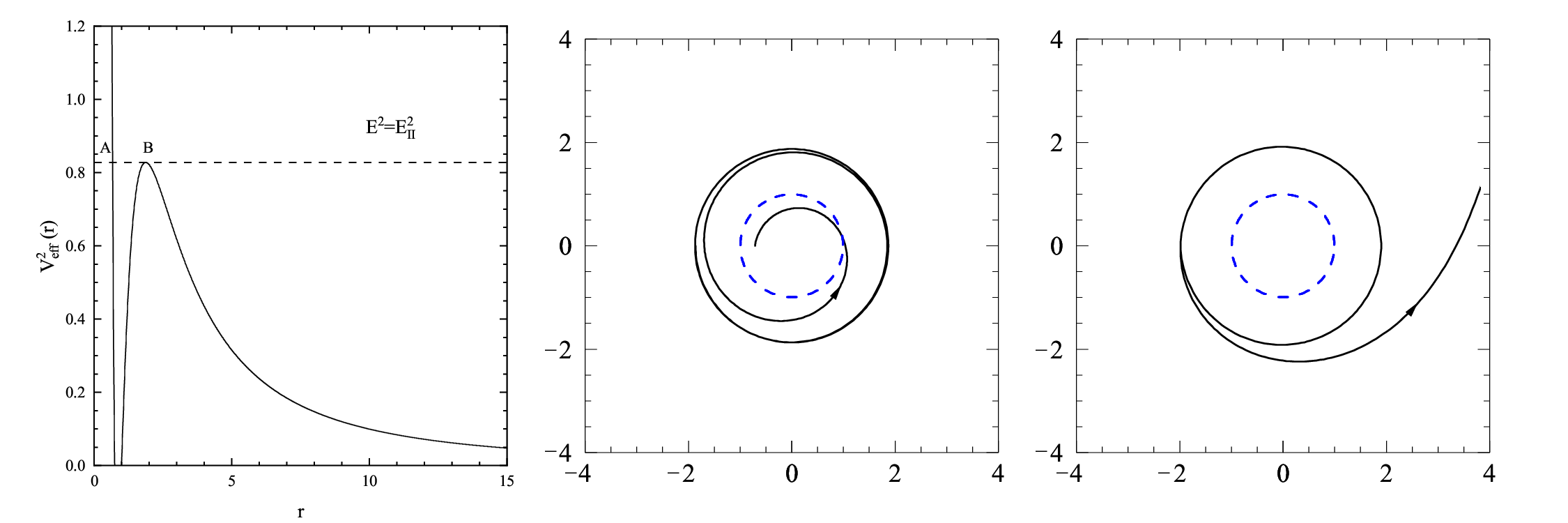}
	\caption{ The effective potential of null bound geodesics and the unstable null circle geodesics in the Bronnikov spacetime with $r_0=0.55$, $L=3.5$, $M=1$, $E_{\uppercase\expandafter{\romannumeral2}}^2=0.83$  }
\end{figure}

\begin{figure}[tbp]
	\includegraphics[width=13cm,height=13cm]{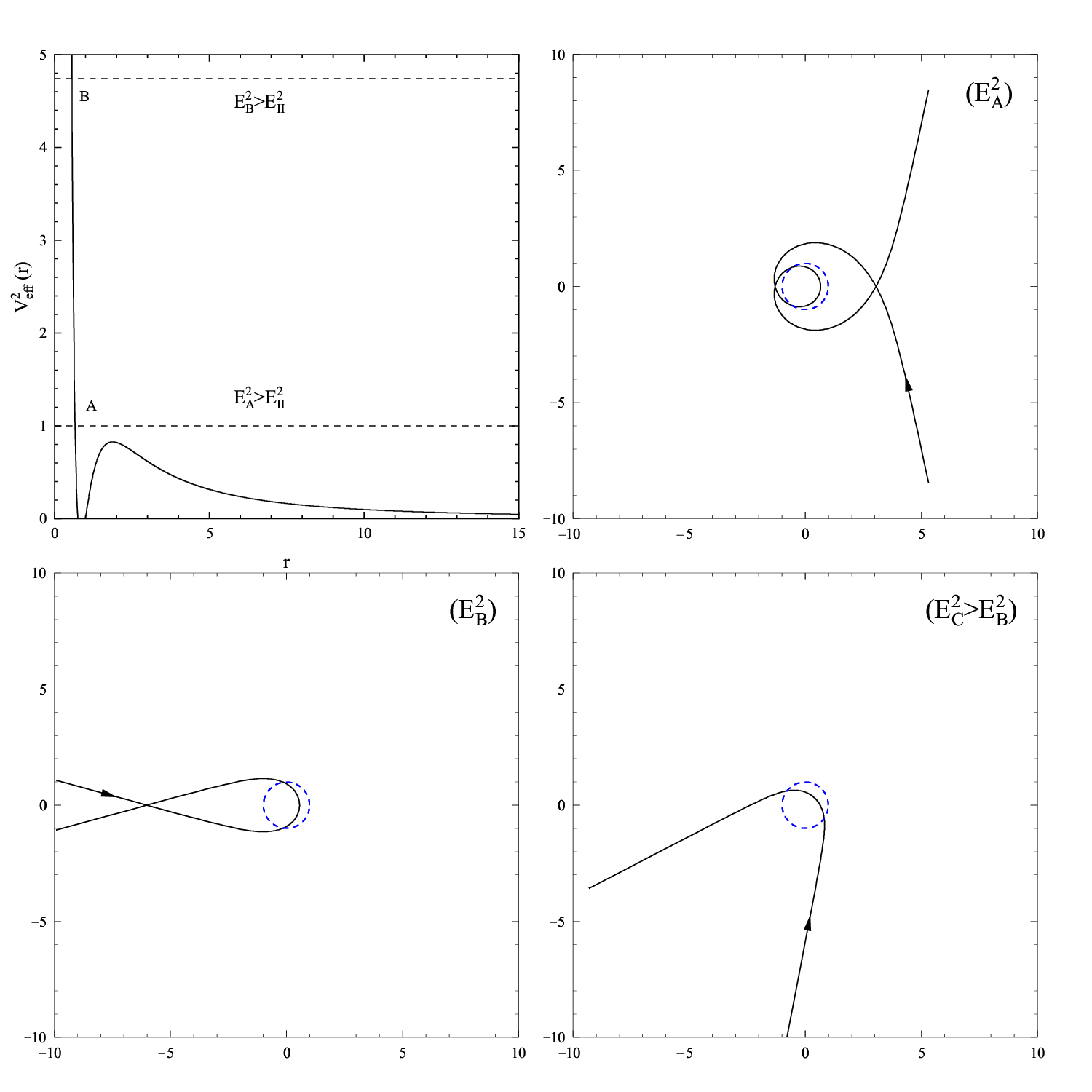}
	\caption{ The null escape geodesics in the Bronnikov spacetime with $r_0=0.55$, $L=3.5$, $M=1$ , $E_{\uppercase\expandafter{\romannumeral2}}^2=0.83$, $E_A^2=1.0$ , $E_B^2=4.7$  }
\end{figure}

\begin{figure}[tbp]
	\includegraphics[width=13cm,height=6cm]{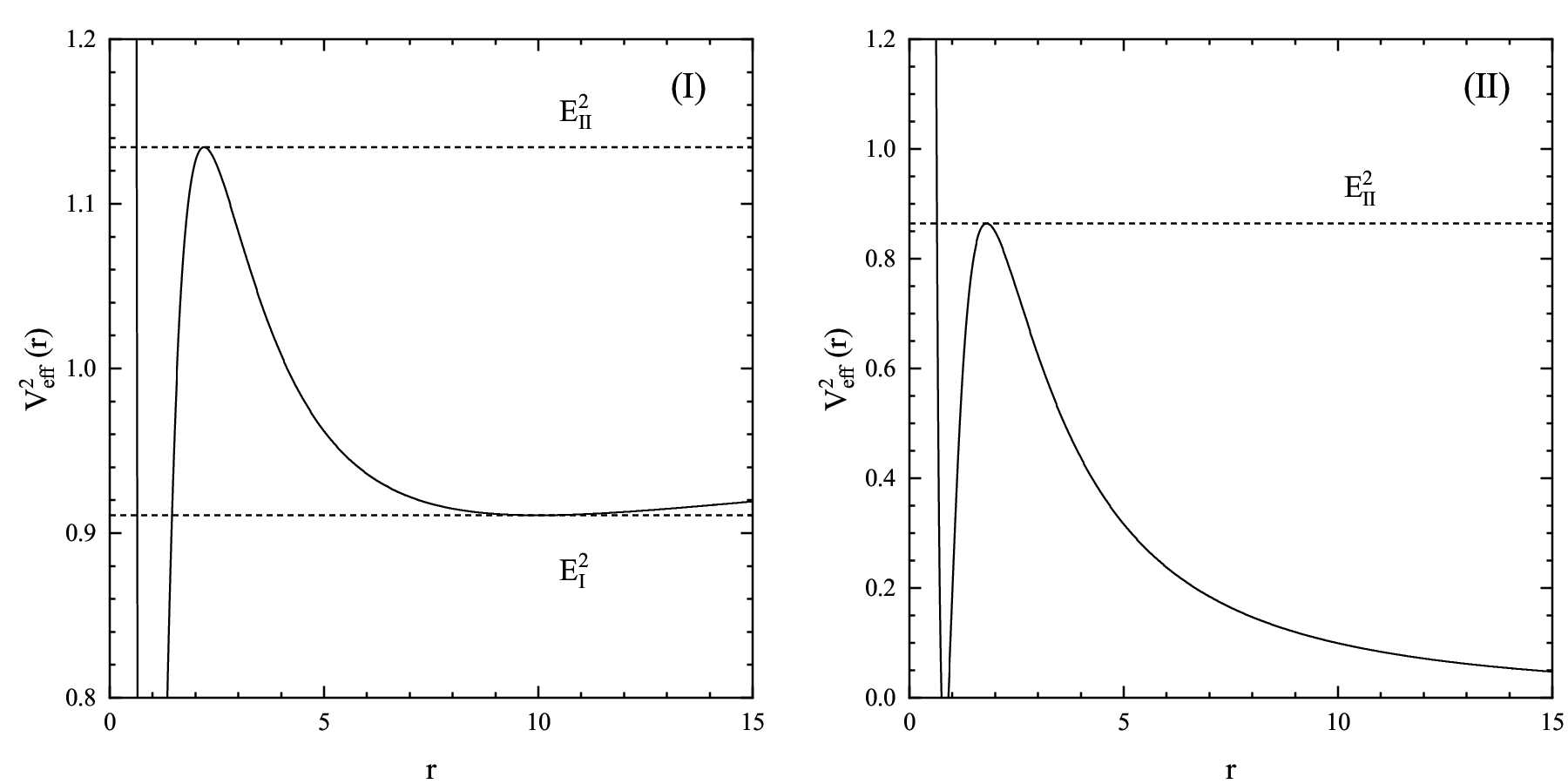}
	\caption{ The effective potential of timelike bound geodesics and null bound geodesics in the Dymnikova spacetime with $r_0=0.45$, $L=3.5$ , $M=1$ ,(\uppercase\expandafter{\romannumeral1}):$E_{\uppercase\expandafter{\romannumeral1}}^2=0.91$,$E_{\uppercase\expandafter{\romannumeral2}}^2=1.13$, (\uppercase\expandafter{\romannumeral2}):$E_{\uppercase\expandafter{\romannumeral2}}^2=0.86$  }
\end{figure}

\begin{figure}[tbp]
	\includegraphics[width=10.6cm,height=15cm]{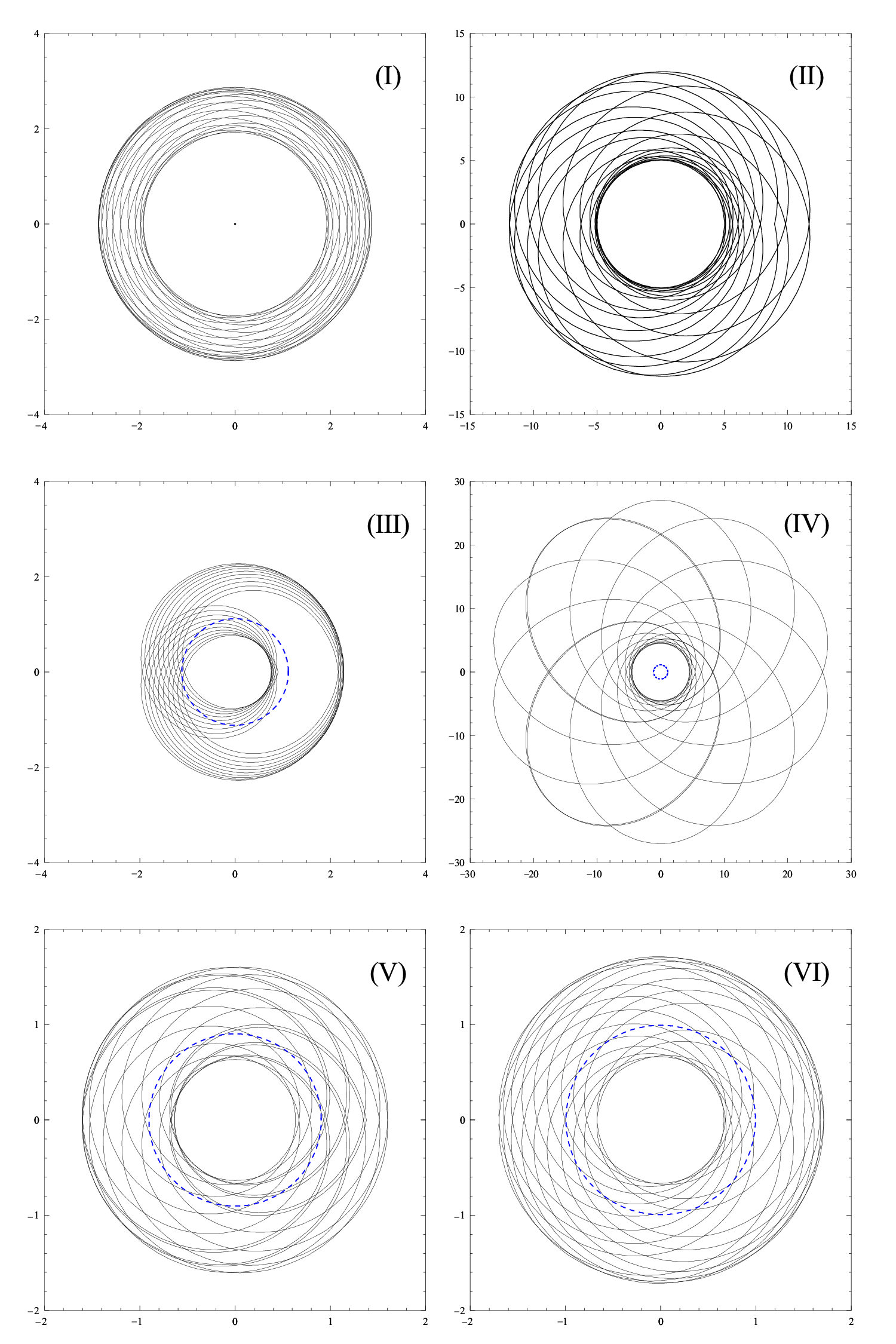}
	\caption{ (\uppercase\expandafter{\romannumeral1}),(\uppercase\expandafter{\romannumeral2}): Two types of timelike bound geodesics in the Hayward spacetime; (\uppercase\expandafter{\romannumeral3}),(\uppercase\expandafter{\romannumeral4}): Two types of timelike bound geodesics in the Ayón-Beato and García spacetime; (\uppercase\expandafter{\romannumeral5}) The timelike bound geodesics in the Bronnikov spacetime (\uppercase\expandafter{\romannumeral6}) The timelike bound geodesics in the Dymnikova spacetime  }
\end{figure}

\begin{figure}[tbp]
	\includegraphics[width=13cm,height=6cm]{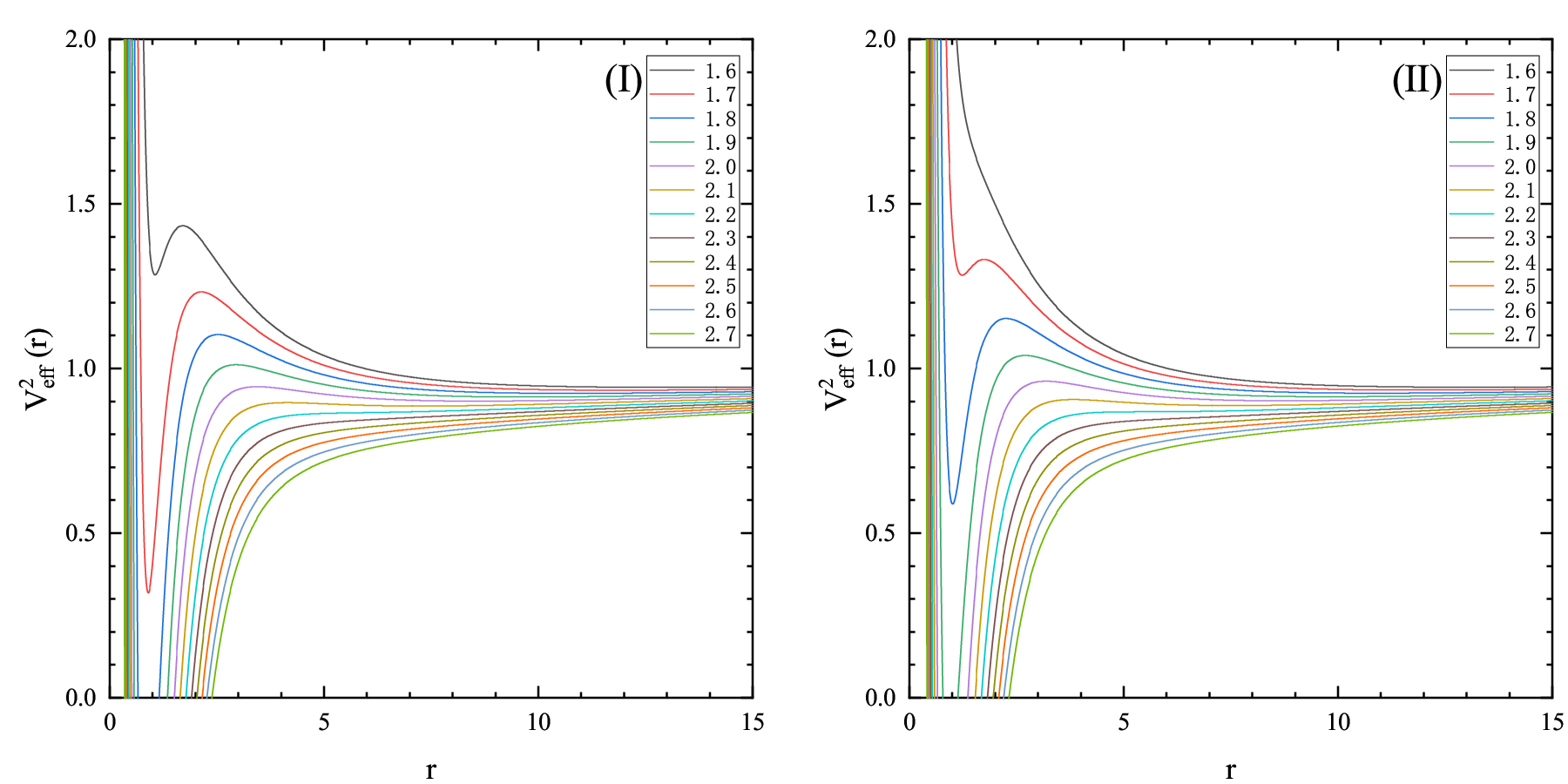}
	\caption{ By fixing the value of $\beta$ and varying the parameter $\alpha$, the plot illustrating the variation of the effective potential is obtained. Here $\eta=1$ , $L=3.5$ , (\uppercase\expandafter{\romannumeral1}):$\beta=0.3$ ,(\uppercase\expandafter{\romannumeral2}) $\beta=0.35$  }
\end{figure}

\begin{figure}[tbp]
	\includegraphics[width=10.6cm,height=20cm]{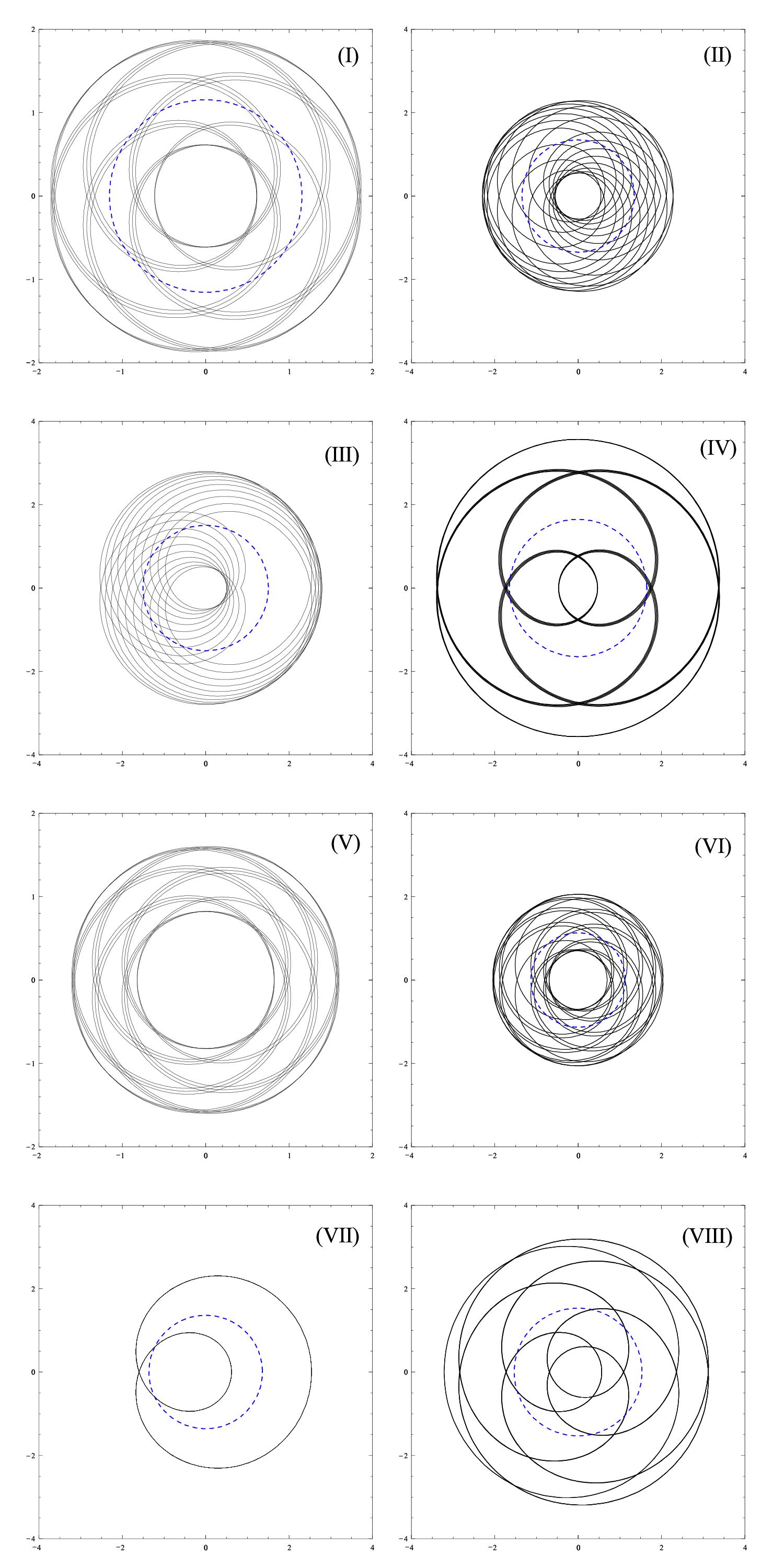}
	\caption{ (\uppercase\expandafter{\romannumeral1})(\uppercase\expandafter{\romannumeral2})(\uppercase\expandafter{\romannumeral3})(\uppercase\expandafter{\romannumeral4})The bound geodesics with continuous variations of $\alpha$ when $\beta$ is fixed at 0.3; (\uppercase\expandafter{\romannumeral5})(\uppercase\expandafter{\romannumeral6})(\uppercase\expandafter{\romannumeral7})(\uppercase\expandafter{\romannumeral8})The bound geodesics with continuous variations of $\alpha$ when $\beta$ is fixed at 0.35  }
\end{figure}


\section{REFERENCES}
\begin{references}
\bibitem{1} N. Heidari, H.Hassanabadi, Physics Letters B 839, 137814 (2023)

\bibitem{2} Chen Wu, Eur. Phys. J. C 78, 283 (2018)

\bibitem{3} B.P. Abbott et al., Phys. Rev. Lett. 116, 241103 (2016)

\bibitem{4} B.P. Abbott et al., Phys. Rev. Lett. 116, 061102 (2016)

\bibitem{5} B.P. Abbott et al., Phys. Rev. Lett. 119, 161101 (2017)

\bibitem{6} Zening Yan, Chen Wu, Wenjun Guo, Nucl. Phys. B 961, 115217 (2020)

\bibitem{7}Cao H. Nam, Gen. Rel. Grav. 50, 57 (2018)

\bibitem{8} M. Umair Shahzad, Abdul Jawad, Farhad Ali, G. Abbas, Chin. J. Phys. 77, 620 (2022)

\bibitem{9} Shobhit Giri, Hemwati Nandan, Gen. Rel. Grav. 53, 76 (2021)

\bibitem{10} Mubasher Jamil, Saqib Hussain, Bushra Majeed, Eur. Phys. J. C 75, 24 (2015)

\bibitem{11} Yen-Kheng Lim, Phys. Rev. D 91, 024048 (2015)

\bibitem{12} Sheng Zhou, Juhua Chen, Yongjiu Wang, Int. J. Mod. Phys. D 21, 1250077 (2012)

\bibitem{13} E. Kapsabelis, P. G. Kevrekidis, P. C. Stavrinos, A. Triantafyllopoulos, Eur. Phys. J. C  82, 1098 (2022)

\bibitem{14} J. Podolsky, Gen. Rel. Grav. 31, 1703 (1999)

\bibitem{15} Chi Zhang, Chen Wu, Gen. Rel. Grav. 50, 18 (2018)

\bibitem{16} Sean A. Hayward, Phys. Rev. Lett. 96, 031103 (2006)

\bibitem{17} J. Bardeen, Paper presented at GR5, Tiflis, U.S.S.R., and published in the conference
proceedings in the U.S.R. (1968)

\bibitem{18} Eloy Ayón-Beato, Alberto García, Phys. Rev. Lett. 80, 5056 (1998)

\bibitem{19} K. A. Bronnikov, Phys. Rev. D 63, 044005 (2001)

\bibitem{20} Waldemar Berej, Jerzy Matyjasek, Dariusz Tryniecki, Mariusz Woronowicz, Gen. Rel. Grav. 38, 885 (2006)

\bibitem{21} Irina Dymnikova, Class. Quant. Grav. 21, 4417 (2004)

\end{references}
\end{document}